\documentclass[conference,10pt]{IEEEtran}
\IEEEoverridecommandlockouts   
\usepackage[utf8]{inputenc}
\usepackage{newtxtext,newtxmath}
\usepackage[dvipsnames]{xcolor}
\usepackage[T1]{fontenc}
\usepackage{array}
\usepackage{url}
\usepackage{balance}
\usepackage{graphicx}
\usepackage{algorithmic}
\usepackage{hyperref}
\usepackage{microtype}
\usepackage{enumitem}
\usepackage{authblk}
\usepackage{listings}
\usepackage{color}
\definecolor{gray}{rgb}{0.5,0.5,0.5}

\ifCLASSOPTIONcompsoc
  \usepackage[caption=false, font=normalsize, labelfont=sf, textfont=sf]{subfig}
  \usepackage[nocompress]{cite}
\else
  \usepackage[caption=false, font=footnotesize]{subfig}
  \usepackage{cite}
\fi

\usepackage[acronym,nohypertypes=acronym]{glossaries}
\usepackage[capitalise]{cleveref}

\setacronymstyle{long-short}

\newcommand{\papertitle}{eSIM Technology in IoT Architecture}

\hypersetup{
  colorlinks=true,
  linkcolor=black,
  urlcolor=blue,
  citecolor=black,
  pdftitle={\papertitle}
}

\newacronym{nist}{NIST}{National Institute of Standards and Technology}
\newacronym{mcv}{MCV}{mission-critical voice}
\newacronym{ui}{UI}{user interface}
\newacronym{nuc}{NUC}{Next Unit of Computing}
\newacronym{ptt}{PTT}{push-to-talk}
\newacronym{rf}{RF}{radio frequency}
\newacronym{dsp}{DSP}{digital signal processing}

\author[]{Hang Yuan}
\author[]{Artiom Baloian}
\author[]{Jan Janak}
\author[]{Henning Schulzrinne}
\affil[]{Department of Computer Science, Columbia University, USA}
\affil[]{Email: hang.yuan@columbia.edu,
ab4659@columbia.edu, janakj@cs.columbia.edu, hgs@cs.columbia.edu}

\begin{document}
\title{\papertitle}
\maketitle

\begin{abstract}
eSIM(embedded SIM) is an advanced alternative to traditional physical SIM cards initially developed by the GSM Association(GSMA) in 2013~\cite{GSMA-eSIM} ~\cite{GSMA-embedded-SIM-arch-1.1}. The eSIM technology has been deployed in many commercial products such as mobile devices. However, the application of the eSIM technology in IoT devices has yet to start being primarily deployed. Understanding the eSIM architecture and the basic ideas of the eSIM provisioning and operations is very important for engineers to promote eSIM technology deployment in more areas, both academics and industries.

The report focuses on the eSIM technology in the IoT architecture and two major operations of Remote SIM Provisioning(RSP) procedure: the Common Mutual Authentication procedure, a process used to authenticate eSIM trusted communication parties over the public internet, and the Profile Downloading procedure, the way to download the Profile from the operator SM-DP+ server and eventually remotely provision the end-user devices.
\end{abstract}

\glsresetall
\section{Introduction}
The traditional SIM card was invented over 30 years ago and has been ubiquitously used~\cite{eSIM-Whitepaper}. People are already familiar with the existence of physical SIM cards, but there are also more challenges with the usage of physical SIM cards. The traditional SIM card is owned and issued by a specific operator. It will be given to the end user as subscription credentials to connect to the operator’s network. To switch the subscription, the end user must obtain a new SIM card issued by the operator and physically swap the SIMs. Considering the management of a large cluster of mobile devices or IoT devices, equipping or managing profiles for this situation could be a disaster where it would need uncountable labor effort to install or switch the physical SIM card. Furthermore, not all the devices support multiple profiles simultaneously. Depending on different device architecture designs, some smartphones reserve only one slot for the physical SIM card to save the device space. The users are limited to choosing specific model of devices with multiple SIM card slots to equip with more than one profile on their devices. Moreover, the physical cards limited the users’ willingness to switch subscriptions between different service providers. To switch the subscription, the customer must order the service and wait for the physical card to arrive to start using the service.

To help resolve the problems of the physical SIM cards and enhance the users’ user experience, eSIM was introduced to achieve remote SIM provisioning. Instead of using the physical SIM card, an embedded SIM (called an eUICC) integrated into the device will accommodate multiple SIM Profiles, which contain the operator and subscriber data to connect to the operator’s network. The end user can scan the QR code received from the operator in turn of the contract to locate and download the new profile. The download and installation operations require the device to be connected to the public network, for example, through the Wi-Fi or a pre-installed Bootstrap Profile. With the Profiles properly installed, the end user can switch between the Profiles to connect their device to whichever operator’s network the end user selects.

The general IoT solution for mobile devices is not applicable to directly apply to IoT devices for the following challenges: (1) compared to the mobile devices which have rich user interface (UI) functionalities for end users to control almost everything, IoT devices are generally lack of UI for end user to directly view or manage the settings; (2) IoT devices are typically deployed in large scale so that designing a reliable and efficient approach to organize a fleet of Iot devices is much more complicate; (3) unlike a mobile device can easily connect to the public networks through Wi-Fi, IoT devices may find difficulties to get internet connections to download profiles from the service provider server. To overcome these challenges, in the eSIM IoT architecture, eSIM IoT Remote Manager (eIM), an additional component, is introduced to enable central management for IoT devices in scale and assist provisioning to IoT devices without public network access.

In this project, guided by Artiom Baloian, I played a pivotal role in gathering and analyzing public technical specifications related to eSIM technology for IoT devices. This experience significantly sharpened my ability to decipher complex technical documentation and offered valuable insights into the innovative and precise solutions employed by industries. Studying industry standards and official documents not only deepened my understanding but also highlighted the nuanced approaches used to address challenges. Jan Janak's feedback during the paper revision phase was instrumental in refining the report for accuracy and coherence. 

This report contains many details of the general eSIM solution for customer devices because the eSIM solution for IoT architecture is developed chiefly based on the prior one. Knowing the architecture and standard procedure for customer devices will help researchers and developers to more easily transplant eSIM technology from current widely deployed customer devices to IoT devices. This report introduces: (1) the architecture of the eSIM, which is simplified, focusing on the major components and each component’s brief introduction; (2) the architecture of the eSIM for IoT environment; (3) The detailed procedure flow of Common Mutual Authentication and the Profile Downloading; (4) the procedure flow of Profile Downloading in IoT architecture.

\section{General eSIM Provisioning Architecture}
The architectures for the eSIM technology have already been specified in GSMA technical specification ~\cite{SGP21}. The architecture involves multiple components for Remote SIM Provisioning (RSP) and the logical interfaces connecting these components. The architecture is organized around four main system elements: the SM-DP+ (Subscription Manager - Data Preparation +), the SM-DS (Subscription Manager - Discovery Server), the LPA (Local Profile Assistant), and the eUICC.

\begin{figure}[htbp]
    \centering
    \includegraphics[scale=0.5]{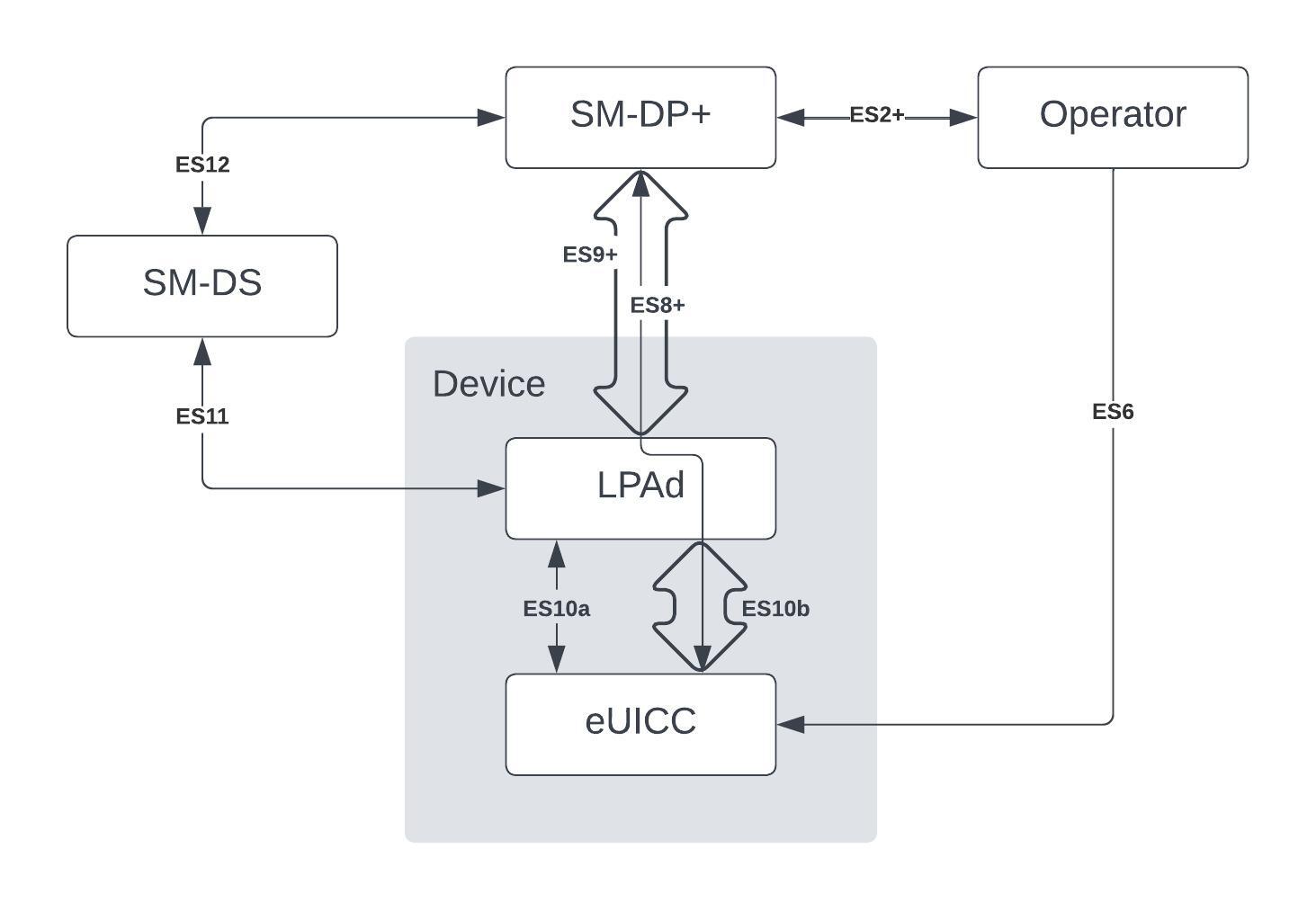}
    \caption{Diagram of Simplified Remote SIM Provisioning(RSP) Architecture for Mobile Devices: major RSP components connected with logical interfaces}
    \label{fig:fig-simplified-arch}
\end{figure}

\subsection{Operator}

Operator means a mobile network operator or mobile virtual network operator. Typically, the operator is a company providing wireless cellular network services and issuing Profiles to its customers to connect to its network.

\subsection{SM-DP+ (Subscription Manager - Data Preparation +)}
The SM-DP+ is responsible for the creation, download, remote management (enable, disable, update, delete), and the protection of Profiles. Typically, SM-DP+ is owned and maintained by the operator to process requests from other secure components and provide a protected Profile package to end users.

\subsection{SM-DS (Subscription Manager - Discovery Server)
}

The SM-DS provides a means for SM-DP+ to reach the device (the eUICC, more specifically) without knowing which network the device is connected to. Since a device can connect to different access networks with different addresses, SM-DP+ can’t connect directly to the device and request certain management operations. The SM-DP+ can post events or send notifications to the SM-DS, wait for the device LPAs to poll the SM-DS when required, and pull the events and notifications. Polling frequency is determined by the eUICC state and by end-user actions.

\subsection{LPA (Local Profile Assistant)}

The Local Profile Assistant (LPA) is a software component that assists the eUICC in executing certain operations. It also supports local management end user interface so they can manage the Profiles on the eUICC. The manufacturer can choose to have LPA in the Device (LPAd) or the eUICC (LPAe). The location of LPA makes a subtle difference to the procedure but doesn’t affect its capability to assist the eUICC. LPA has four primary functions: 1. Local Discovery Service; 2. Local Profile Download; 3. Local User Interface; and 4. LPA Proxy.

\begin{itemize}
    \item Local Discovery Service: LPA is responsible for periodically querying the SM-DS server it subscribed to respond to any event published by the SM-DP+ server.
    \item Local Profile Download: LPA assists to the profile downloading procedure as a proxy role for efficiently downloading a Bound Profile Package. The LPA will first download the bound profile package from the SM-DP+. The profile package will then be transferred to eUICC in segments for later installation. The end user, SM-DP+ server, or discovery service can request this operation.
    \item Local User Interface: For customer devices requiring end-user interactions (e.g., obtaining consent from the end user), customized mobile applications can utilize UI implemented by the LPA.
    \item LPA Proxy: Mobile applications can utilize LPA Proxy service to provide a better user experience, like showing the progress of specific operations.
\end{itemize}

\subsection{eUICC}

The eUICC is a discrete or integrated tamper-resistant component consisting of hardware and software capable of securely hosting applications as well as confidential and cryptographic data such as Profiles and certificates~\cite{SGP01}. The eUICC may accommodate more than one profile depending on the implementation and configuration of the eUICC chip and any restrictions imposed by the mobile network operator or service provider. According to the GSMA standards~\cite{SGP01}, only one Profile should be enabled at anytime. However, enabling multiple profiles is possible with the help of the device’s Operating System and hardware ~\cite{esim-mep}.

\subsection{Logical Interface}

Multiple logical interfaces are implemented for communication purposes between any two of the RSP components. The logical secure elements establish a logical connection with each other. These interfaces are responsible for transiting the data for specific operations and enforcing the security requirements for communication. Each interface may support many operations for particular purposes. For example, the ES8+ interface is a logical interface that provides a secure end-to-end channel between the SM-DP+ and the eUICC for the administration and the associated Profile during download and installation~\cite{SGP31}. In ES8+ interface, ES8+.InitialiseSecureChannel is one of the operations that ES8+ interface should support.

\subsection{eSIM on Android Device}

Since Android 9, the eSIM feature has been supported at the operating system level~\cite{esim-overview}. Specific OS-layer modules have been added to enable the eSIM feature and assist with the customized Carrier App and LPA. More support features are added to Android as the version goes further. Starting with Android 10, Android can support devices with multiple eSIMs. Beginning in Android 13, enabling multiple profiles at the same time becomes available~\cite{esim-mep}.

\begin{figure}[htbp]
    \centering
    \includegraphics[scale=0.5]{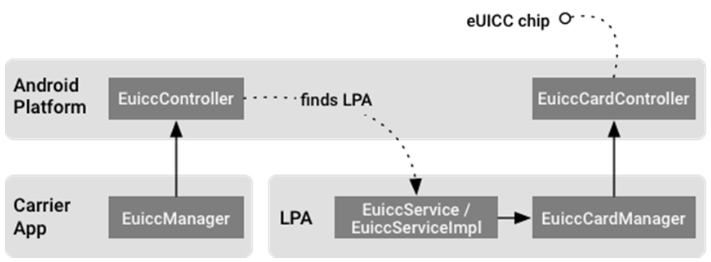}
    \caption{Android eSIM Module Architecture: operations are executed through different system and customized modules}
    \label{fig:android-esim}
\end{figure}

The diagram in~\cref{fig:android-esim} shows the simplified architecture of Android regarding the eSIM implementation~\cite{esim-euicccardmanager}. To unify the interfaces across different services, Android has published a set of interfaces for developers to develop their own LPA or Carrier App. These interfaces include the EuiccService/EuiccServiceImpl to create customized LPA, and the EuiccManager to implement customized Carrier App. Android also implemented functionalities in the OS layer to support eSIM working, including EuiccController to register the LPA instance, EuiccCardController to communicate with the eUICC chip, and the EuiccCardManager to assist customized Linteractionact with OS-layer EuiccCardController.

The LPA is a standalone system app that should be included in the Android build image. Management of the profiles on the eSIM is generally done by the LPA, as it serves as a bridge between the SM-DP+ (remote service that prepares, stores, and delivers profile packages to devices) and the eUICC chip. The LPA APK can optionally include a UI component, called the LPA UI or LUI, to provide a central place for the end user to manage all embedded subscription profiles. The Android framework automatically discovers and connects to the best available LPA and routes all the eUICC operations through an LPA instance.

\section{IoT eSIM Provisioning Architecture}
Unlike mobile devices, IoT devices may not be equipped with a user interface, or may only have limited network connectivity. To support such devices, an additional component – eSIM IoT remote Manager (eIM) is introduced to the architecture to assist in managing a cluster of IoT devices. The architecture of eSIM for IoT device is introduced in GSMA's eSIM IoT architecture specification~\cite{SGP31}.

\begin{figure}[htbp]
    \centering
    \includegraphics[scale=0.5]{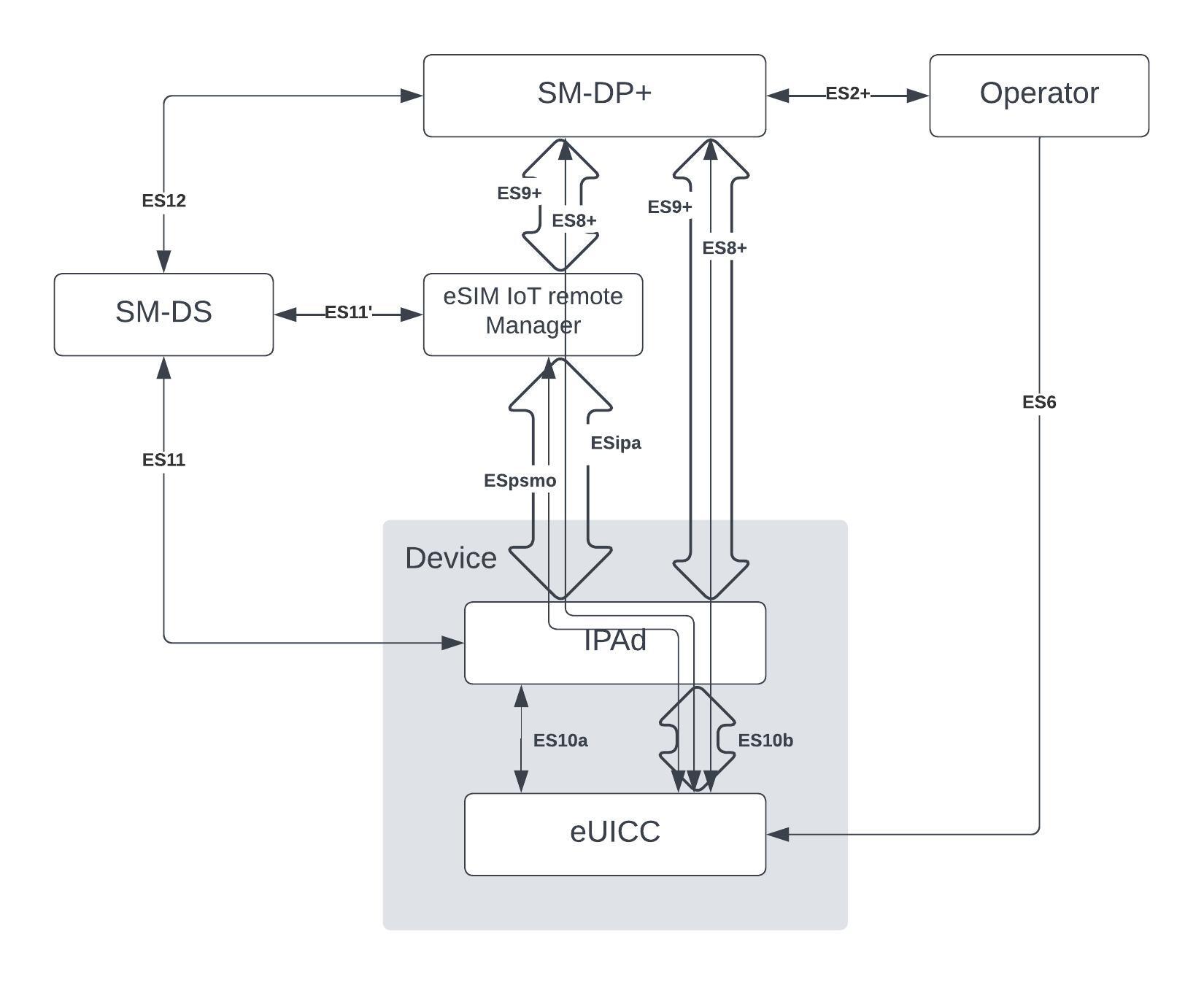}
    \caption{Diagram of the eIM and corresponding logical interfaces are added to the general RSP Architecture for IoT Devices}
    \label{fig:simplifed-arch-IoT}
\end{figure}

\subsection{eIM (eSIM IoT remote Manager)}

The eIM is responsible for remote Profile State Management Operations (PSMO) on a single IoT device or a fleet of IoT devices. The eIM can communicate with the SM-DP+ and SM-DS on behalf of the device to exchange data as a proxy so that the device can overcome its UI or network connectivity constraints during Profile management. The eIM can either be a stand-alone component or a component of a higher-level functional system (e.g., a device management platform).

\subsection{IPA (IoT Profile Assistant)}
Similar to the LPA in the customer device, IPA also provides functions that enable the eUICC in the IoT device to be provisioned by the SM-DP+. To better support the IoT environment, the IPA replaces some features that the LPA provides with unique ones for IoT devices specifically. The primary functionalities are: 1. Discovery Service; 2. Profile Download; 3. PSMO Conveying; 4. Notification Handling. The Discovery Service and Profile Download functions are identical to the LPA’s Local Discovery Service and Local Profile Download. The PSMO Conveying helps convey the PSMO (Profile State Management Operation) and related results between eIM and eUICC. The IPA is also responsible for forwarding notifications to the eIM and/or the SM-DP+.
IPA can also be located in the Device (IPAd) or the eUICC (IPAe).

\section{eSIM remote Provisioning for mobile devices}

This section introduces the provisioning details in general architecture (for mobile devices) and two crucial procedures: Common Mutual Authentication and Profile Downloading. These two procedures are required to complete the remote provisioning by downloading the profile from the remote SM-DP server to the local device and finally install it into the eUICC. Understanding the procedures in general architecture will help understand the procedures for eSIM IoT architecture.

\subsection{Communication}

The communication amoung the RSP components mainly depends on the TCP/IP connection with HTTPS protocol to exchange operation requests and responses with required data. In addition to network data protection, GSMA also requires that the certificate must sign all the data transmitted to ensure data integrity.
The description of some data objects in this specification is based on ASN.1 (Abstract Syntax Notation One). This provides a flexible description of those data objects ~\cite{SGP22}.

\subsection{Certificate}

All the certificates used in the eSIM provisioning are X.509 certificates specifically. ~\cite{SGP22}

Two types of certificates are used in the provisioning stage: GSMA CI/partnered Sub-root CI signed certificates for Common Mutual Authentication, and any public root CA (including GSMA CI) signed certificates as TLS certificates to support HTTPS connection. The certificate mentioned below is limited to the certificate that can be chained back to GSMA CI unless explicitly.

Certificates are generated and distributed in the provisioning stage. The certificate chain must have the GSMA CI or its partner root CA as the final root CA. These include SM-DP+, SM-DS, and EUM certificates distributed by GSMA or its partner sub-root CA. The eUICC certificate is signed and installed by its manufacturer.

\begin{figure}[htbp]
    \centering
    \includegraphics[width=\linewidth]{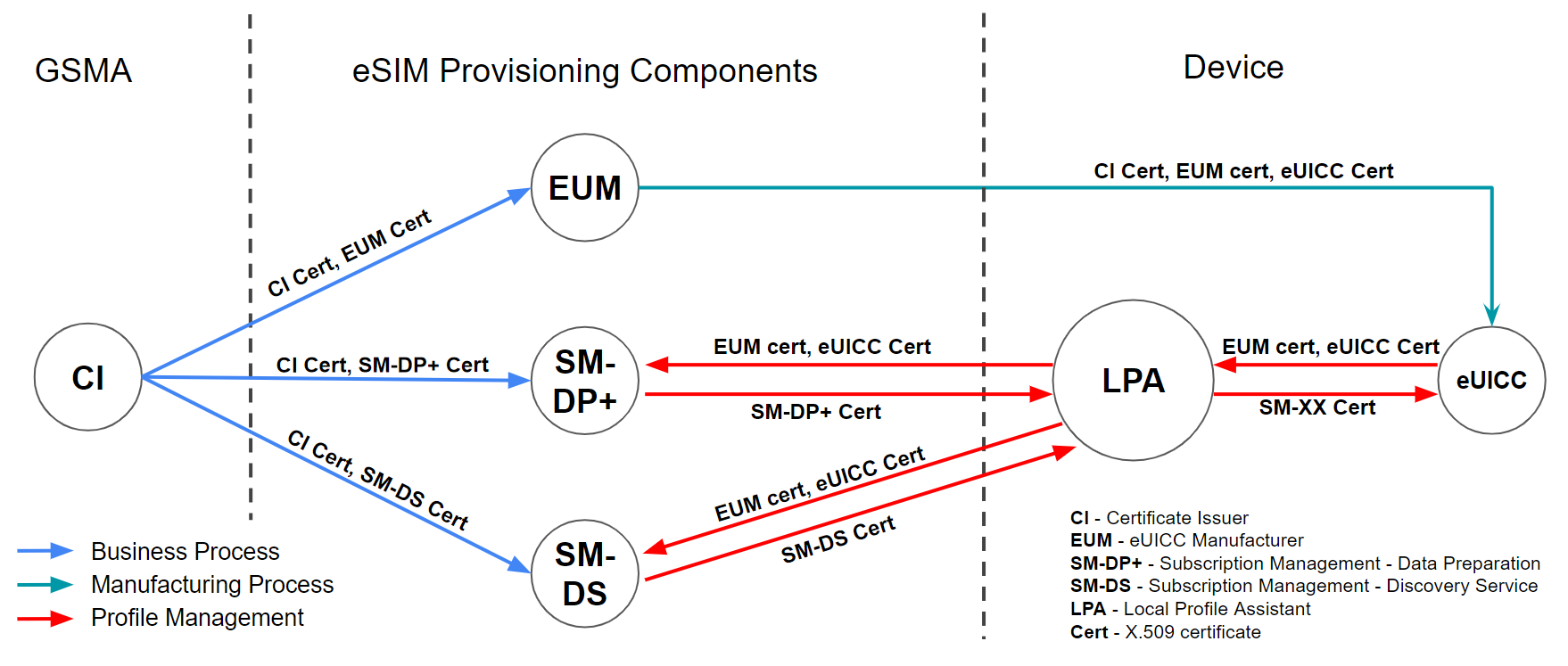}
    \caption{Certificate Exchange: the certificate is issued and chained by different level of issuers in different processes}
    \label{fig:certificate-exchange}
\end{figure}

The diagram~\cref{fig:certificate-exchange} shows the certificate exchange between the RSP components. The certificates are distributed in three processes: Business Process, manufacturing process, and profile management. The certificate issuer is also responsible for reissuing or updating the certificates if required.

\subsubsection{Certificate Validation}
The certificate can be considered invalid if any following condition fails:
\begin{itemize}
    \item It has a valid signature.
    \item It is signed by a GSMA CI, or an independent eSIM CA, or a trusted chain of certificates up or a GSMA CI or an independent eSIM CA.
    \item It has not been revoked, and no certificate in the trust chain has been revoked.
    \item It has not expired.
\end{itemize}

\subsubsection{Certificate Revocation}

All the certificates shall be revokable, and the certificate's signer is responsible for revoking the certificates it has signed. The eUICC, LPA/IPA, SM-DS, and SM-DP+ shall have knowledge of revoked public key certificates.

In case the certificate of any entities (SM-DP+, SM-DS, EUM) is compromised (e.g., private key theft), the eSIM CA(s) shall revoke the compromised certificate. The revoked certificate will lead to all the certificates signed by this certificate in the trust chain becoming revoked as well.

\subsection{Common Mutual Authentication}

Common Mutual Authentication~\cite{SGP22} is a crucial part of the security appliance during the communication between any two RSP components. It ensures that both communicating parties verify each other’s identity with a valid certificate with a trust chain to GSMA CA. The authentication procedure gets executed right after the TLS session is established and before any operational data is exchanged. Any connection must be immediately terminated if the authentication fails at any time.

\begin{figure*}[htbp]
    \centering
    \includegraphics[width=\linewidth]{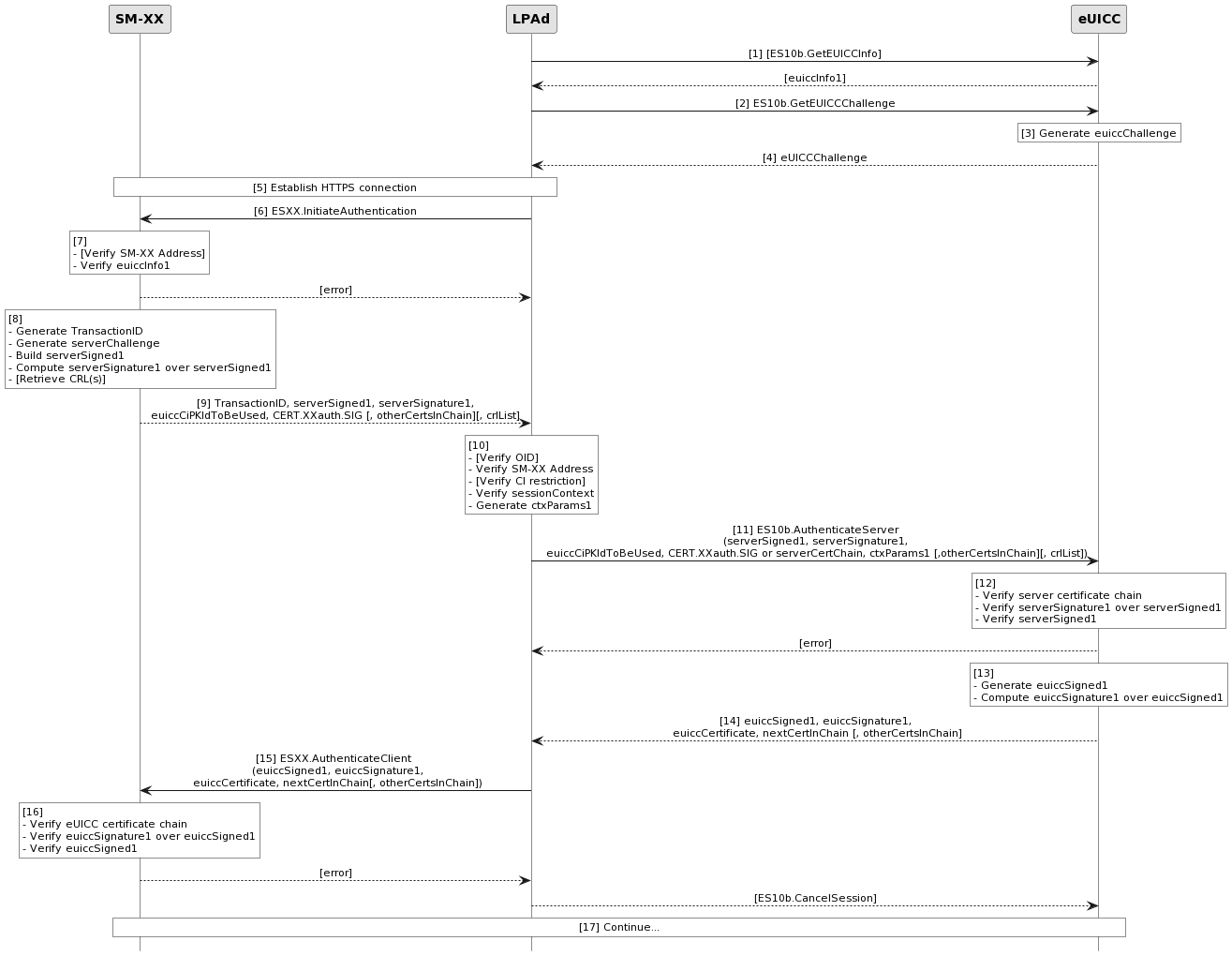}
    \caption{Procedure Flow of Common Mutual Authentication}
    \label{fig:common-mutual-auth}
\end{figure*}

The diagram in~\cref{fig:common-mutual-auth} contains the steps and certificate exchange during the authentication process. The SM-XX (SM-DP+ or SM-DS) will initially offer the certificate to eUICC to verify, and the eUICC will give its own certificate in return for mutual authentication. Once the authentication is completed successfully, the components can continue using the connection for other operational procedures. Otherwise, the component with invalid certificates will be considered unauthenticated, and the connection will be abandoned.

The description of some data objects is based on ASN.1~\cite{SGP22} in the specifications. For the readability of the report, some data objects in ASN.1 format are replaced with simple lists of data elements.

For simplicity, only the key data elements will be mentioned for data object.

Start condition:
\begin{enumerate}
    \item[a. ] The SM-XX is provisioned with its Certificate(s), its private key(s), the eSIM CA RootCA Certificate(s), its TLS Certificate(s), its TLS Private Key(s), and the SM-XX SubCA Certificates, if any, in the trust chains of the certificate(s) mentioned above.\\
    The eUICC is also provisioned with its Certificate(s), its private key(s), the EUM Certificate(s), the eSIM CA RootCA Public Key(s), and the SubCA Certificate(s), if any, in the trust chains.
\end{enumerate}

Procedure:
\begin{enumerate}
    \item[1. ] Optionally, the LPA MAY request eUICC Information euiccInfo1 from eUICC by calling the \texttt{ES10b.GetEUICCInfo} function. This is required if the LPAd hasn’t already retrieved this information. 
    
        With the returned \texttt{euiccInfo1}, if there is a restriction of the allowed eSIM CA RootCA public key(s), the LPA SHALL create a new instance of \texttt{euiccInfo1} by removing all PK identifiers that do not match the given identifier. If there is no available identifier, stop the procedure.
        
        The LPA sends \texttt{GetEuiccInfo1Request} to the eUICC.
        
        The eUICC responds with \texttt{EUICCInfo1} data object, containing:

        \begin{itemize}
            \item \texttt{euiccCiPKIdListForVerification} -- List of eSIM CA RootCA Public Key Identifiers supported on the eUICC for signature verification. Each Public Key Identifier refers to an eSIM CA RootCA Certificate, which implicitly defines other cryptographic algorithms which should be used in this RSP session.
            \item \texttt{euiccCiPKIdListForSigning} -- List of eSIM CA RootCA Public Key Identifier supported on the eUICC for signature creation that can be verified by a certificate chain.
            \item \texttt{euiccCiPKIdListForSigningV3} [OPTIONAL, SupportedFromV3.0.0] -- List of eSIM CA RootCA Public Key Identifiers supported on the eUICC for signature creation that can be verified by a certificate chain different.
            \item \texttt{euiccRspCapability} [OPTIONAL, MandatoryFromV3.0.0]
        \end{itemize}
    
    \item[2. ] The LPA requests an eUICC Challenge from the calling the \texttt{ES10b.GetEUICCChallenge} function.
    
        The LPA sends \texttt{GetEuiccChallengeRequest} to the eUICC.
    
    \item[3. ] The eUICC SHALL generate a random eUICC Challenge in Octet16 which SHALL be signed later by the SM-XX for SM-XX authentication by the eUICC.
    
    \item[4. ] The eUICC returns the eUICC Challenge to the LPA.

    \item[5. ] The LPA establishes a new HTTPS connection with the the SM-XX in server authentication mode. The TLS session establishment SHALL perform a new key exchange.
    
    \item[6. ] The LPA SHALL call the \texttt{ESXX.InitiateAuthentication} function. Below is an example of a request with mock data:
\begin{lstlisting}
HTTP POST <Server InitiateAuthentication Endpoint> HTTP/1.1
Host: <Server Address>
User-Agent: <User Agent>
X-Admin-Protocol: gsma/rsp/v<x.y.z>
Content-Type: application/json;charset=UTF-8
Content-Length: XXX
{
    "euiccChallenge":"ZVVpY2NDaGFsbGVu...",
    "euiccInfo1": "RmVHRnRjR3hsUW1G...",
    "sm[xx]Address": "smxx.example.com",
    "lpaRspCapability": "ODAwMjAzRjg="
}
\end{lstlisting}

    \item[7. ] The SM-XX SHALL verify 
        \begin{enumerate}
            \item[a. ] the SM-XX Address sent by the LPAd is valid. If the SMXX Address is not valid, the SM-XX SHALL return an error.
            \item[b. ] the list of eSIM CA RootCA Public Keys that are associated to the eUICC credentials (\texttt{euiccCiPKIdListForSigning} and \texttt{euiccCiPKIdListForSigningV3} if present, as contained in \texttt{euiccInfo1}).
            \item[c. ] the received eSIM CA RootCA Public Key Identifier list (\texttt{euiccCiPKIdListForVerification} contained in the \texttt{euiccInfo1}). If it cannot provide a \texttt{CERT.XXauth.SIG} which chain can be verified by an eSIM CA RootCA Public Key supported by the eUICC:
                \begin{itemize}
                    \item If the LPAd RSP capabilities indicated \texttt{euiccCiUpdateSupport}, the SM-XX SHOULD select its preferred \texttt{CERT.XXauth.SIG}.
                    \item In all other cases, it SHALL return an error.
                \end{itemize}
        \end{enumerate}
        If the LPAd receives an error in this step, then the LPA SHALL stop the procedure.

    \item[8. ] The SM-XX SHALL perform the following:
        \begin{itemize}
            \item Generate a \texttt{TransactionID} which is used to uniquely identify the RSP session and to correlate the multiple ESXX request messages that belong to the same RSP session.
            \item Generate an SM-XX Challenge (\texttt{serverChallenge}) which SHALL be signed later by the eUICC for the eUICC authentication.
            \item Select one eSIM CA RootCA Public Key among those provided within \texttt{euiccCiPKIdListForSigning} or \texttt{euiccCiPKIdListForSigningV3}, that is supported by the RSP Server for signature verification and indicate it in \texttt{euiccCiPKIdToBeUsed} or in \texttt{euiccCiPKIdToBeUsedV3} respectively.
            \item Generate a \texttt{serverSigned1} data structure.
            \item Compute the \texttt{serverSignature1} over \texttt{serverSigned1} using the \texttt{SK.XXauth.SIG} corresponding to the \texttt{CERT.XXauth.SIG} determined in step 7.
            \item If both eUICC and LPA indicate \texttt{crlStaplingV3Support}, retrieve the latest CRL for each Certificate in the chain that has a \texttt{cRLDistributionPoints} extension set (unless it is already available).
        \end{itemize}
    
    \item[9. ] The SM-XX SHALL return to the LPA the response. Below is an example of the response with mock data:
\begin{lstlisting}
HTTP/1.1 200 OK
X-Admin-Protocol: gsma/rsp/v<x.y.z>
Content-Type: application/json;charset=UTF-8
Content-Length: XXX
{
    "header": {
    "functionExecutionStatus": {
    "status": "Executed-Success"
}
},
    "transactionId": "0123456789ABCDEF",
    "serverSigned1": "VGhpcyBpcyBub3QgY...",
    "serverSignature1": "RKNFZsbFVUa0...",
    "euiccCiPKIdToBeUsed": " BBQAAQIDBAUg...",
    "serverCertificate": "RUU2NTQ0ODQ5N...",
    "otherCertsInChain": ["q83vASM..."]
}
\end{lstlisting}

    \item[10. ] The LPAd SHALL:
        \begin{itemize}
            \item If the SM-XX is an SM-DP+ and if its \texttt{OID} was provided earlier, verify the \texttt{OID} as specified in the procedure where this call flow is used.
            \item Verify that the SM-XX Address returned by the SM-XX matches the SM-XX Address that the LPA has provided in step (6).
            \item If there is a restriction to a single allowed eSIM CA RootCA public key identifier, verify that the Subject Key Identifier of the eSIM RootCA corresponding to \texttt{CERT.XXauth.SIG} matches this value.
            \item If the LPAd indicated \texttt{euiccCiUpdateSupport}, verify that the Subject Key Identifier of the Root Certificate corresponding to \texttt{CERT.XXauth.SIG} is included in \texttt{euiccInfo1.euiccCiPKIdListForVerification}. If the verification fails, the LPAd SHALL inform the End User and stop the procedure, after which it MAY perform the eUICC Root Public Key update procedure (not introduced in this report) indicating that Subject Key Identifier.
            \item (Optional) Verify that each Certificate in the chain and each CRL in the list (if present) is valid with respect to its time window, i.e., notBefore and thisUpdate are in the past, and notAfter and nextUpdate are in the future, with regard to the current time known by the Device.
            \item If any verification fails, the LPA SHALL inform the End User and stop the procedure.
            \item Generate a data structure, ctxParams1, to be given to the eUICC to be included in signed data.
        \end{itemize}

    \item[11. ] The LPAd SHALL call \texttt{ES10b.AuthenticateServer} function to send \texttt{AuthenticateServerRequest} with the following data:

    \begin{itemize}
        \item \texttt{serverSigned1}  -- Signed information
        \item \texttt{serverSignature1} 
        \item \texttt{euiccCiPKIdToBeUsed} [OPTIONAL] -- eSIM CA RootCA Public Key Identifier to be used; MAY also have zero length
        \item \texttt{serverCertificate} -- RSP Server Certificate CERT.XXauth.SIG 
        \item \texttt{ctxParams1} -- session context parameters including some flags that certain operations need to be processed
        \item \texttt{otherCertsInChain} [OPTIONAL, SupportedFromV3.0.0] -- CertificateChain, the remaining part of the CERT.XXauth.SIG certificate chain (if any) 
        \item \texttt{crlList} [OPTIONAL, SupportedFromV3.0.0] -- SEQUENCE OF CertificateList
    \end{itemize}

    \item[12. ] The eUICC SHALL:
        \begin{itemize}
            \item Verify the \texttt{CERT.XXauth.SIG} and other certificates in the chain, if any, starting with \texttt{CERT.XXauth.SIG}, using the relevant \texttt{PK.CI.SIG}.
            \item Verify the \texttt{serverSignature1} performed over \texttt{serverSigned1}.
            \item Verify that \texttt{euiccChallenge} contained in \texttt{serverSigned1} matches the one generated by the eUICC during step (3).
            \item Verify that the eSIM CA RootCA Public Key Identifier indicated in either \texttt{euiccCiPKIdToBeUsed} or \texttt{euiccCiPKIdToBeUsedV3} is supported and related credentials are available for signing.
            \item If the \texttt{sessionContext} indicates \texttt{crlStaplingV3Used}, verify the validity of each CRL, and verify that no Certificate in the chain is revoked.
        \end{itemize}

        If all the verification succeed, the SM-XX is authenticated by the eUICC.

        If any verification fails, the eUICC SHALL return the following error status and the procedure SHALL stop.

\begin{lstlisting}
-- ASN1START
AuthenticateResponseError ::= SEQUENCE {
    transactionId TransactionId,
    authenticateErrorCode AuthenticateErrorCode
}
AuthenticateErrorCode ::= INTEGER {
    invalidCertificate(1), 
    invalidSignature(2), 
    unsupportedCurve(3), 
    noSession(4), 
    invalidOid(5), 
    euiccChallengeMismatch(6), 
    ciPKUnknown(7), 
    transactionIdError(8), 
    missingCrl(9), 
    invalidCrlSignature(10), 
    revokedCert(11), 
    invalidCertOrCrlTime(12), 
    invalidCertOrCrlConfiguration(13), 
    invalidIccid(14), 
    undefinedError(127)
}
-- Note: ErrorCode (8)-(14) are SupportedFromV3.0.0
-- ASN1STOP    
\end{lstlisting}

    \item[13. ] The eUICC SHALL:
        \begin{itemize}
            \item Generate the \texttt{euiccSigned1} data structure.
            \item Compute the \texttt{euiccSignature1} over \texttt{euiccSigned1} using \texttt{SK.EUICC.SIG}. When generating the \texttt{euiccSignature1}, the eUICC SHALL use the credentials identified in the previous step.
        \end{itemize}

        The \texttt{euiccSigned1} should be encoded as follows:

\begin{lstlisting}
-- ASN1START
EuiccSigned1 ::= SEQUENCE {
    transactionId TransactionId,
    serverAddress UTF8String, -- The RSP Server address as an FQDN 
    serverChallenge Octet16, -- The RSP Server Challenge 
    euiccInfo2 EUICCInfo2, 
    ctxParams1 CtxParams1
}
-- ASN1STOP    
\end{lstlisting}

    \item[14. ] The eUICC SHALL return the coded response data to the LPA.

\begin{lstlisting}
-- ASN1START
AuthenticateServerResponse ::= CHOICE {
    authenticateResponseOk AuthenticateResponseOk,
    authenticateResponseError AuthenticateResponseError
}
AuthenticateResponseOk ::= SEQUENCE {
    euiccSigned1 EuiccSigned1, -- Signed information
    euiccSignature1 OCTET STRING, --EUICC_Sign1
    euiccCertificate Certificate, -- eUICC Certificate (CERT.EUICC.SIG)
    nextCertInChain Certificate, -- The Certificate certifying the eUICC Certificate
    otherCertsInChain CertificateChain [OPTIONAL, SupportedFromV3.0.0] -- Other Certificates in the eUICC certificate chain, if any
}
-- ASN1STOP    
\end{lstlisting}

    \item[15. ] The LPAd SHALL call the "ESXX.AuthenticateClient" function with input data comprising euiccSigned1, euiccSignature1 and the eUICC certificate chain.

\begin{lstlisting}
HTTP POST <Server Authentication Endpoint>HTTP/1.1
Host: <Server Address>
User-Agent: <User Agent>
X-Admin-Protocol: gsma/rsp/v<x.y.z>
Content-Type: application/json;charset=UTF-8
Content-Length: XXX
{
    "serverSigned1": "VGhpcyBpcyBub3QgYSByZWFsIHZhbHVl",
    "serverSignature1": "RKNFZsbFVUa05qUm14e",
    "euiccCiPKIdToBeUsed": " BBQAAQIDBAUGBwgJCgsMDQ4PEBESEw==",
    "serverCertificate": "RUU2NTQ0ODQ5NDA0RlpSRUZERA==...",
    "euiccSigned1":"VGhpcyBpcyBub3QgYSByZWFsIHZhbHVl",
    "euiccSignature1": "RKNFZsbFVUa05qUm14e",
    "euiccCertificate": "eSEgWgLNwgWEkC5ttEYL6HWkMw7H...",
    "nextCertInChain": "9nRxyFfBsJ2WmsPX8mXAM9CvuYAB...",
    "otherCertsInChain": ["q83vASM..."]
}   
\end{lstlisting}

    \item[16. ] On reception of the \texttt{ESXX.AuthenticateClient} function call, the SM-XX SHALL:
        \begin{itemize}
            \item Correlate it with the \texttt{ESXX.InitiateAuthentication} function processed in steps (7) and (8), by verifying the two \texttt{TransactionIDs} match.
            \item Verify that the Root Certificate of the eUICC certificate chain corresponds to the \texttt{euiccCiPKIdToBeUsed} or \texttt{euiccCiPKIdToBeUsedV3} that the SM-XX selected when executing the \texttt{ESXX.InitiateAuthentication} function.
            \item Verify that the eUICC Certificate chain is valid.
            \item Verify the \texttt{euiccSignature1} performed over \texttt{euiccSigned1} using the \texttt{PK.EUICC.SIG} contained in the \texttt{CERT.EUICC.SIG}.
            \item Verify that \texttt{serverChallenge} contained in \texttt{euiccSigned1} matches the one generated by the SM-XX during step (7).
            \item Verify that the eUICC and LPA RSP capabilities match those received in \texttt{ESXX.InitiateAuthentication}.
        \end{itemize}

        If any verification fails, the SM-XX SHALL return a relevant error status to the LPAd.

        If all verification succeeds, the SM-XX SHALL return a response comprising the pending RSP operation to the LPAd depending on the procedure within which this procedure is used.

        If the LPAd receives an error status, or only the \texttt{TransactionID} from the SM-DP+ in this step, then the LPAd SHALL send \texttt{ES10b.CancelSession} to the eUICC with a reason \texttt{sessionAborted}.

\begin{lstlisting}
-- ASN1START
CancelSessionRequest ::= SEQUENCE {
    transactionId TransactionId, -- The TransactionID generated by the RSP Server 
    reason CancelSessionReason
}
CancelSessionReason ::= INTEGER {
    endUserRejection(0),
    postponed(1),
    timeout(2),
    pprNotAllowed(3),
    metadataMismatch(4),
    loadBppExecutionError(5),
    sessionAborted(16)
    enterpriseProfilesNotSupported(17)
    enterpriseRulesNotAllowed(18)
    enterpriseProfileNotAllowed(19)
    enterpriseOidMismatch(20)
    enterpriseRulesError(21)
    enterpriseProfilesOnly(22)
    lprNotSupported(23)
    lprNetworkDataNotAllowed(24)
    emptyProfileOrSpName(25)
    rpmDisabled(27)
    invalidRpmPackage(28)
    loadRpmPackageError(29)
    undefinedReason(127)
}
--Note: CancelSessionReason (16)-(29) are SupportedForRpmV3.0.0
-- ASN1STOP    
\end{lstlisting}

         A response will be returned by eUICC to the LPA regarding the \texttt{CancelSession} request.
\begin{lstlisting}
-- ASN1START
CancelSessionResponse ::= CHOICE {
    cancelSessionResponseOk CancelSessionResponseOk,
    cancelSessionResponseError INTEGER {invalidTransactionId(5), undefinedError(127)}
} 
CancelSessionResponseOk ::= SEQUENCE { 
    euiccCancelSessionSigned EuiccCancelSessionSigned, -- Signed information 
    euiccCancelSessionSignature
} 
EuiccCancelSessionSigned ::= SEQUENCE { 
    transactionId TransactionId, 
    smdpOid OBJECT IDENTIFIER, -- SM-DP+ OID as contained in CERT.DPauth.SIG 
    reason CancelSessionReason 
} 
-- ASN1STOP
\end{lstlisting}

    \item[17. ] This common call flow SHALL be followed by additional steps depending on the procedure within which it is used.
\end{enumerate}

End Condition:
\begin{enumerate}
    \item[a. ] Both SM-XX and LPA authenticated each other and are able to maintain the current session for the other procedures.
\end{enumerate}

\subsection{Profile Downloading}

    The GSMA technical specifications require the LPA to support at least one of the following profile-downloading approaches:
    
    \begin{itemize}
        \item Profile Download from default SM-DP+
        \item Profile Download with Activation Code
        \item Profile Download via SM-DS
    \end{itemize}

    The most common way of downloading profiles for customer devices is downloading profiles with an activation code offered by the operator and received by the end user in return for the contract made with the operator. The diagram below is the general procedure for downloading the profile from the SM-DP+.

\begin{figure*}[htbp]
    \centering
    \includegraphics[width=\linewidth]{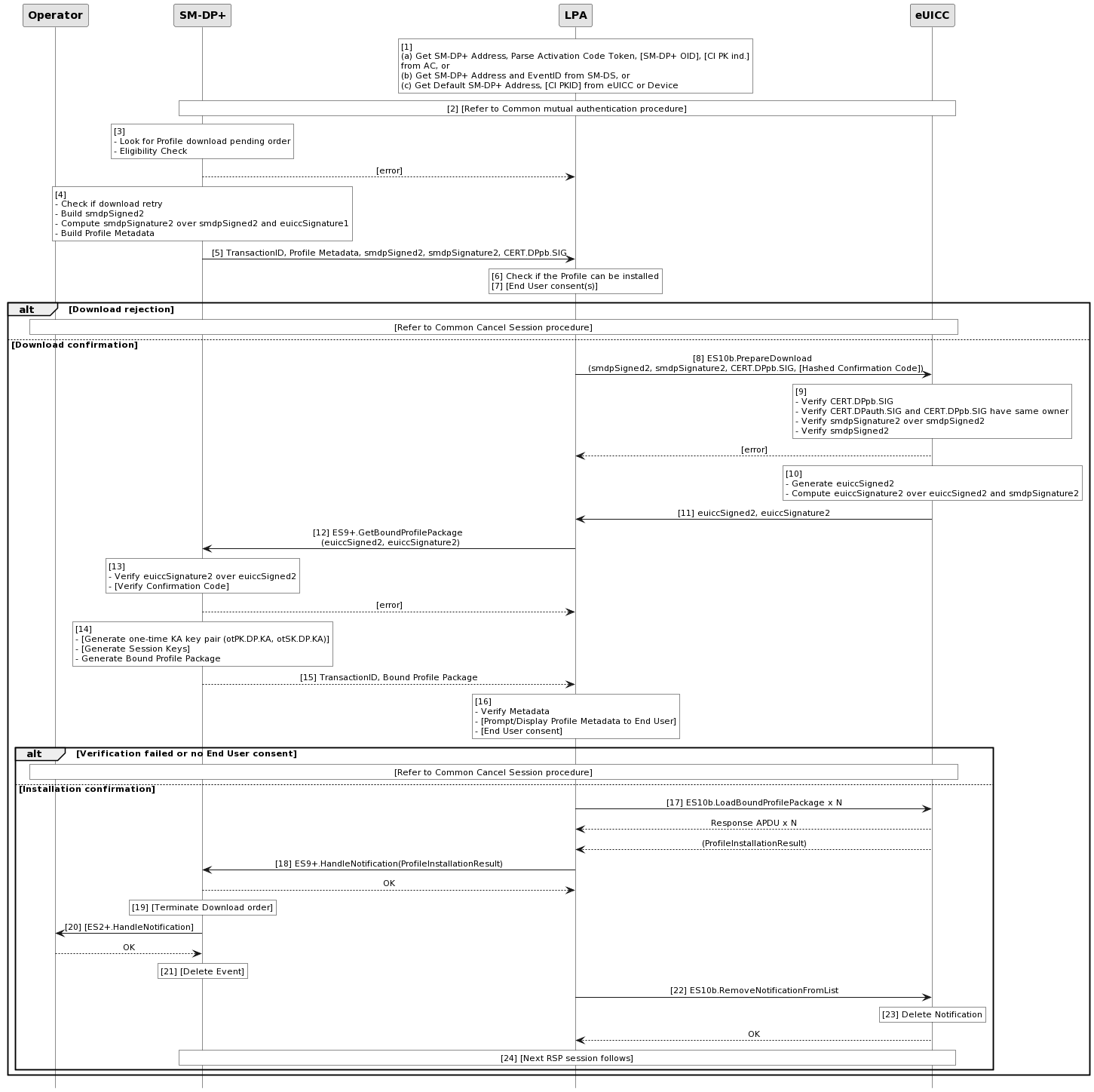}
    \caption{General Procedure of Profile Download}
    \label{fig:profile-download}
\end{figure*}

Start condition:
\begin{enumerate}
    \item[a. ] The ordering process of the Profile has been completed, and Activation Code is generated by the Operator and made available to the end user.
        The end user shall make an order for an eSIM service subscription from the service provider and the corresponding Operator of the service provider will process the order and generate an activation code and distribute to the end user as part of the contract.
        The activation code is a string with multiple sections of important information to target the specific SM-DP+ RSP server address and ordered subscription identifier. To avoid end users manually inputting the activation code, the Operator can further create a QR code using the activation code and the end users can scan the QR code.
\end{enumerate}

Procedure:
\begin{enumerate}
    \item[1. ] The LPA obtains the SM-DP+ address and relevant needed data binding to the order of the operator.
        \begin{enumerate}
            \item[a. ] If the end user receives the QR code, the LPA is responsible to parse all the required information from the QR code or rely on the end user to input the information.
            \item[b. ] The SM-DP+ address and eventID can be retrieved from the SM-DS.
            \item[c. ] The Default SM-DP+ Address is configured and is ready to download an available order from the SM-DP+.
        \end{enumerate}

    \item[2. ] The Common Mutual Authentication is processed successfully. All the required information for the later use, including \texttt{TransactionID}, certificates and cert chains to be used, \texttt{euiccInfo1}, \texttt{euiccInfo2}, and \texttt{ctxParams1} have been exchanged in this phase.
    
    \item[3. ] The SM-DP+ SHALL:
        \begin{itemize}
            \item Verify that there is a related pending Profile download order for the provided \texttt{MatchingID}.
            \item If this Profile download order is already linked to an \texttt{EID}, verify that it matches the \texttt{EID} of the authenticated eUICC.
            \item Verify that the Profile corresponding to the pending Profile download order is in 'Released' state, or, in case of a retry due to a previous installation failure, in 'Downloaded' state
        \end{itemize}

        If any of these verifications fail, the SM-DP+ SHALL return a relevant error status and the procedure SHALL stop. Otherwise, the SM-DP+ SHALL:
        \begin{itemize}
            \item Increment the count of download attempts for the identified Profile. If the maximum number of attempts has been exceeded, the SM-DP+ SHALL terminate the corresponding Profile download order and notify the Operator by calling the \texttt{ES2+.HandleNotification} function with the \texttt{notificationEventStatus} indicating 'Failed' with the relevant error status, and the procedure SHALL stop.
            \item Perform appropriate eligibility checks based on \texttt{deviceInfo} and \texttt{euiccInfo2} retrieved from the \texttt{ctxParams1} in step (2).
        \end{itemize}

        If the eligibility check fails, the SM-DP+ SHALL:
        \begin{itemize}
            \item Set the Profile corresponding with the pending Profile download order in 'Error' state.
            \item Return an error status to the LPAd and the procedure SHALL stop.
        \end{itemize}

    \item[4. ] The SM-DP+ SHALL:
        \begin{itemize}
            \item Determine if a Confirmation Code is required for this pending order.
            \item Determine whether the Profile is already bound to the \texttt{EID} from a previous unsuccessful download attempt. If so, the SM-DP+ MAY indicate in its response the \texttt{otPK.EUICC.KA} it wants to use.
            \item Determine if an RPM Package for the \texttt{EID} is also pending.
            \item Generate an \texttt{smdpSigned2} data structure containing associated data elements.
            \item Compute the \texttt{smdpSignature2}.
            \item Generate the Profile Metadata of the Profile.
        \end{itemize}

    \item[5. ] The SM-DP+ returns \texttt{ES9+.AuthenticateClient} response to the LPA.

\begin{lstlisting}
--ASN1START
AuthenticateClientResponseEs9 ::= CHOICE {
    authenticateClientOk AuthenticateClientOk,
    authenticateClientError INTEGER {
        eumCertificateInvalid(1),
        eumCertificateExpired(2),
        euiccCertificateInvalid(3),
        euiccCertificateExpired(4),
        euiccSignatureInvalid(5),
        matchingIdRefused(6),
        eidMismatch(7),
        noEligibleProfile(8),
        ciPKUnknown(9),
        invalidTransactionId(10),
        insufficientMemory(11),
        ciPKMismatch(12)
        euiccRspCapabilityHasChanged(13)
        lpaRspCapabilityHasChanged(14)
        deviceChangeNotSupported(15)
        deviceChangeNotAllowed(16)
        iccidUnkwon(17)
        invalidInputData(124)
        missingInputData(125)
        functionProviderBusy(126)
        undefinedError(127)
    },
    authenticateClientOkRpm AuthenticateClientOkRpm,
    authenticateClientOkDeviceChange AuthenticateClientOkDeviceChange
}
--Note: authenticateClientOkRpm, authenticateClientOkDeviceChange, and authenticateClientError(12)-(17) and (124)-(126) are SupportedFromV3.0.0
AuthenticateClientOk ::= SEQUENCE {
    transactionId  TransactionId,
    profileMetadata StoreMetadataRequest [OPTIONAL],
    smdpSigned2 SmdpSigned2 OPTIONAL, -- Signed information
    smdpSignature2 OCTET STRING [OPTIONAL],
    smdpCertificate Certificate -- CERT.DPpb.SIG
}
AuthenticateClientOkRpm ::= SEQUENCE {
    transactionId TransactionId,
    smdpSigned3 SmdpSigned3,
    smdpSignature3 OCTET STRING
}
AuthenticateClientOkDeviceChange ::= SEQUENCE {
    transactionId TransactionId,
    smdpSigned4 SmdpSigned4, -- Signed information
    smdpSignature4 OCTET STRING
    serviceProviderMessageForDc LocalisedTextMessage [OPTIONAL], -- Service Provider Message For Device Change
}
--ASN1STOP
\end{lstlisting}

    \item[6. ]On reception of the SM-DP+ response, the LPAd SHALL check if the Profile can be installed as described hereunder. For this check, the LPAd MAY use previously fetched Rules Authorization Table and/or list of installed Profiles. If the LPAd has not already fetched the required information, the LPAd SHALL request those from the eUICC by calling the \texttt{ES10b.GetRAT} and/or \texttt{ES10c.GetProfilesInfo} functions.

    (For simplicity, the relevant rules to the enterprise have been excluded)

    \begin{itemize}
        \item If the Profile Metadata contains PPR(s), the LPAd SHALL check if the PPR(s) are allowed based on the Rules Authorization Table. If one or more PPR(s) are not allowed, the LPAd SHALL perform the Common Cancel Session procedure with reason \texttt{pprNotAllowed}.
        \item If the Profile Metadata contains PPR1, and an Operational Profile is installed, the LPAd SHALL perform the Common Cancel Session procedure with reason \texttt{pprNotAllowed}.
        \item If the Profile Metadata contains an LPR Configuration and the Device or the eUICC does not support the LPR, the LPAd SHOULD perform the Common Cancel Session procedure with reason \texttt{lprNotSupported}.
        \item If the Profile Metadata contains an empty string profileName and/or serviceProviderName, the LPAd MAY perform the Common Cancel Session procedure with reason emptyProfileOrSpName if \texttt{cancelForEmptySpnPnSupport} is supported by both the SM-DP+ and the eUICC or with reason undefinedReason otherwise.
    \end{itemize}

    \item[7. ] According to certain Profile Policy Rule(s) or Enterprise Rule(s), the LPA may ask the end user for different confirmation to consent the downloading process.

    \item[8. ] The LPAd SHALL call the \texttt{ES10b.PrepareDownload} function optionally including the Hashed Confirmation Code.

        The command data SHALL be coded as follows:

\begin{lstlisting}
-- ASN1START
PrepareDownloadRequest ::= SEQUENCE {
    smdpSigned2 SmdpSigned2, -- Signed information
    smdpSignature2 OCTET STRING
    hashCc [OPTIONAL], -- Hash of confirmation code
    smdpCertificate Certificate, -- CERT.DPpb.SIG
}
SmdpSigned2 ::= SEQUENCE {
    transactionId TransactionId -- The TransactionID generated by the SM-DP+
    ccRequiredFlag BOOLEAN, -- Indicates if the Confirmation Code is required
    bppEuiccOtpk  OCTET STRING OPTIONAL, -- otPK.EUICC.KA already used for binding the BPP
    rpmPending NULL [OPTIONAL, SupportedForRpmV3.0.0]
}
-- ASN1STOP
\end{lstlisting}

    \item[9. ] The eUICC SHALL:
        \begin{itemize}
            \item Verify that \texttt{CERT.DPpb.SIG} is valid.
            \item Verify that \texttt{CERT.DPauth.SIG} and \texttt{CERT.DPpb.SIG} belong to the same entity and are certified by the same certificate.
            \item Verify \texttt{smdpSignature2}.
            \item Verify that the \texttt{TransactionID} contained in \texttt{smdpSigned2} matches the \texttt{TransactionID} of the on-going RSP session.
        \end{itemize}

        If any of the verifications fail, the eUICC SHALL return a relevant error status and the procedure SHALL stop.

        The response data SHALL be coded as follows.

\begin{lstlisting}
-- ASN1START
PrepareDownloadResponse ::= CHOICE {
    downloadResponseOk PrepareDownloadResponseOk,
    downloadResponseError PrepareDownloadResponseError
}
PrepareDownloadResponseError ::= SEQUENCE {
    transactionId TransactionId,
    downloadErrorCode DownloadErrorCode
}
DownloadErrorCode ::= INTEGER {invalidCertificate(1), invalidSignature(2), noSession(4), invalidTransactionId(5), undefinedError(127)}
-- ASN1STOP
\end{lstlisting}

        In case of the error \texttt{invalidTransactionId}, the \texttt{transactionId} in the \texttt{PrepareDownloadResponse} SHALL be set to the value from the \texttt{AuthenticateServerRequest}.

    \item[10. ] The eUICC SHALL:
        \begin{itemize}
            \item Use the one-time key pair associated with the \texttt{bppEuiccOtpk} if it is provided by the SM-DP+ and it is still stored in the eUICC, or generate a new one-time key pair.
            \item Generate the \texttt{euiccSigned2} data structure.
            \item Compute the \texttt{euiccSignature2}.
        \end{itemize}

    \item[11. ] The eUICC should send the response to LPA coded as follows.

\begin{lstlisting}
-- ASN1START
PrepareDownloadResponse ::= CHOICE {
    downloadResponseOk PrepareDownloadResponseOk,
    downloadResponseError PrepareDownloadResponseError
}
PrepareDownloadResponseOk ::= SEQUENCE {
    euiccSigned2 EUICCSigned2, -- Signed information
    euiccSignature2 OCTET STRING
}
EUICCSigned2 ::= SEQUENCE {
    transactionId TransactionId,
    euiccOtpk OCTET STRING, -- otPK.EUICC.KA
    hashCc Octet32 [OPTIONAL], -- Hash of confirmation code
}
-- ASN1STOP
\end{lstlisting}

    \item[12. ] The LPAd calls the \texttt{ES9+.GetBoundProfilePackage} function.
\begin{lstlisting}
--ASN1START
GetBoundProfilePackageRequest ::= SEQUENCE {
    transactionId TransactionId,
    prepareDownloadResponse PrepareDownloadResponse
}
--ASN1STOP
\end{lstlisting}

    \item[13. ] The SM-DP+ SHALL verify the \texttt{euiccSignature2}.

        If a Confirmation Code is required, the SM-DP+ SHALL verify it.

        If any verification in step 6 or 7 failed, the SM-DP+ SHALL return an error status to the LPAd and the procedure SHALL stop.

    \item[14. ] The SM-DP+ SHALL perform the following: Dependent on whether a re-usable BPP is present and whether the eUICC can accept it, the SM-DP+ SHALL do one of the following:
        \begin{itemize}
            \item Reuse the BPP
            \item Rebind the BPP
            \item Return an error
            \item Create a new BPP
        \end{itemize}

    \item[15. ] The SM-DP+ returns \texttt{ES9+.GetBoundProfilePackage} response and set the Profile corresponding to the Profile download order in 'Downloaded' state.

\begin{lstlisting}
--ASN1START
GetBoundProfilePackageResponse ::= CHOICE {
    getBoundProfilePackageOk GetBoundProfilePackageOk,
    getBoundProfilePackageError INTEGER {
        euiccSignatureInvalid(1),
        confirmationCodeMissing(2),
        confirmationCodeRefused(3),
        confirmationCodeRetriesExceeded(4),
        bppRebindingRefused(5),
        downloadOrderExpired(6),
        invalidTransactionId(95),
        invalidInputData(124),
        missingInputData(125),
        functionProviderBusy(126),
        undefinedError(127)
    }
    --Note: getBoundProfilePackageError(124)-(126) are SupportedFromV3.0.0
}
GetBoundProfilePackageOk ::= SEQUENCE {
    transactionId TransactionId,
    boundProfilePackage BoundProfilePackage
}
--ASN1STOP
\end{lstlisting}

    \item[16. ] The LPAd SHALL perform additional processing using the Profile Metadata contained within the Bound Profile Package:
        \begin{itemize}
            \item If the LPAd previously used the Profile Metadata returned by \texttt{ES9+.AuthenticateClient} (i.e., in step (5)) then
                \begin{itemize}
                    \item The LPAd SHALL verify that the Profile Policy Rules and the Enterprise Configuration have not changed. If this verification fails, the LPAd SHALL execute the Common Cancel Session procedure with reason \texttt{pprNotAllowed} or \texttt{metadataMismatch}.
                    \item The LPAd SHOULD verify that all other Profile Metadata elements it used in that earlier step (such as the Profile Name, Icon, etc.) have not changed. If the verification fails, the LPAd MAY inform the End User and offer the End User to postpone or reject the Profile installation. Alternatively, the LPAd MAY stop the download by executing the Common Cancel Session procedure with reason \texttt{metadataMismatch}.
                \end{itemize}
            \item If the LPAd has not previously captured the End User consent(s) related to the Profile download as defined in step (7), it SHALL do so at this point as described in that step.
        \end{itemize}

    \item[17. ] The LPA repeatedly calls the \texttt{ES10b.LoadBoundProfilePackage} function containing the four ES8+ functions to the eUICC. These ES8+ functions should be processed one by one through the ES10b channel. The ES8+ functions wrapped in the ES10b function are:
        \begin{enumerate}
            \item[a. ] \texttt{ES8+.InitialiseSecureChannel}
            \item[b. ] \texttt{ES8+.ConfigureISDP}
            \item[c. ] \texttt{ES8+.StoreMetadata}
            \item[d. ] \texttt{ES8+.LoadProfileElements}
        \end{enumerate}

    The \texttt{ES10b.LoadBoundProfilePackage} function can be repeatedly called if the data being transferred is too large. The oversize data will be sent in fragments.

    If all the Profile Elements are successfully processed and installed, with or without any warning, the last response of the \texttt{ES10b.LoadBoundProfilePackage} function SHALL deliver the signed Profile Installation Result as defined below.

\begin{lstlisting}
-- ASN1START
-- Definition of Profile Installation Result
ProfileInstallationResult ::= SEQUENCE {
    profileInstallationResultData ProfileInstallationResultData,
    euiccSignPIR EuiccSign
}
ProfileInstallationResultData ::= SEQUENCE {
    transactionId[0] TransactionId, -- The TransactionID generated by the SM-DP+
    notificationMetadata NotificationMetadata,
    smdpOid OBJECT IDENTIFIER, -- SM-DP+ OID (value from CERT.DPpb.SIG)
    finalResult CHOICE {
        successResult SuccessResult,
        errorResult ErrorResult
    }
}

EuiccSign ::= OCTET STRING -- eUICC's signature

SuccessResult ::= SEQUENCE {
    aid OCTET STRING (SIZE (5..16)), -- AID of ISD-P
    ppiResponse OCTET STRING -- contains (multiple) 'EUICCResponse' of the Profile Package Interpreter as defined in [5]
}

ErrorResult ::= SEQUENCE {
    bppCommandId BppCommandId,
    errorReason ErrorReason,
    ppiResponse OCTET STRING OPTIONAL -- contains (multiple) 'EUICCResponse' of the Profile Package Interpreter as defined in [5]
}

BppCommandId ::= INTEGER {
    initialiseSecureChannel(0),
    configureISDP(1),
    storeMetadata(2),
    storeMetadata2(3),
    replaceSessionKeys(4),
    loadProfileElements(5)
}

ErrorReason ::= INTEGER {
    incorrectInputValues(1),
    invalidSignature(2),
    invalidTransactionId(3),
    unsupportedCrtValues(4),
    unsupportedRemoteOperationType(5),
    unsupportedProfileClass(6),
    bspStructureError(7),
    bspSecurityError(8),
    installFailedDueToIccidAlreadyExistsOnEuicc(9),
    installFailedDueToInsufficientMemoryForProfile(10),
    installFailedDueToInterruption(11),
    installFailedDueToPEProcessingError (12),
    installFailedDueToDataMismatch(13),
    testProfileInstallFailedDueToInvalidNaaKey(14),
    pprNotAllowed(15),
    enterpriseProfilesNotSupported(17),
    enterpriseRulesNotAllowed(18),
    enterpriseProfileNotAllowed(19),
    enterpriseOidMismatch(20),
    enterpriseRulesError(21),
    enterpriseProfilesOnly(22),
    lprNotSupported(23),
    unknownTlvInMetadata(26),
    installFailedDueToUnknownError(127)
}
--Note: 
-- ASN1STOP ErrorReason(17)-(26) are SupportedFromV3.0.0
\end{lstlisting}

    Installation notifications as configured in the Profile Metadata, if any, SHALL be generated.

    \item[18. ] The LPA calls the \texttt{ES9+.HandleNotification} function in order to deliver the Profile Installation Result to the SM-DP+. The SM-DP+ acknowledges the reception of the Notification to the LPA.

    \item[19. ] The SM-DP+ SHALL:
        \begin{itemize}
            \item Retrieve the pending download order identified by the \texttt{TransactionID}. If \texttt{TransactionID} is unknown, the SM-DP+ SHALL terminate its processing.
            \item (Conditional) Terminate the pending download order and set the corresponding Profile in state 'Installed' or 'Error' (section 3.1.6) as indicated by the Profile Installation Result.
        \end{itemize}
        
    \item[20. ] (Conditional) The SM-DP+ SHALL call the \texttt{ES2+.HandleNotification} with:
        \begin{itemize}
            \item \texttt{notificationEvent} indicating 'BPP installation';
            \item \texttt{notificationEventStatus} reflecting the value received in \texttt{ES9+.HandleNotification};
            \item \texttt{notificationReceiverIdentifier} reflecting the \texttt{functionRequesterIdentifier} value of the associated \texttt{ES2+.ConfirmOrder};
            \item \texttt{notificationIdentifier} reflecting the \texttt{functionCallIdentifier} value of the associated \texttt{ES2+.ConfirmOrder};
        \end{itemize}

    \item[21. ] (Conditional) If this procedure is executed in the context of option (b), the SM-DP+ SHALL execute the SM-DS event deletion procedure (section 3.6.3).

    \item[22. ] On reception of the acknowledgement message from the SM-DP+ the LPAd SHALL call \texttt{ES10b.RemoveNotificationFromList} with the corresponding seqNumber.

    \item[23. ] The eUICC SHALL delete the Profile Installation Result from its non-volatile memory.

    \item[24. ] (Conditional) If the LPAd has received rpmPending in the response of \texttt{ES9+.AuthenticateClient} function call, the LPAd SHOULD initiate an additional RSP Session with the SM-DP+, setting the operationType to indicate rpm. If this RSP session was triggered by an Event Record from an SM-DS, the pending RSP session with the SM-DP+ SHOULD be executed before continuing processing any remaining Event Records from that SM-DS.
\end{enumerate}
\section{eSIM remote Provisioning for IoT devices}

Compared to the mobile devices, IoT devices are lack of UI for end users to manage the profiles and it's hard to manage a cluster of IoT devices in scale. This section introduces the involvement of eSIM IoT remote Manger (eIM) in the provisioning procedure to overcome the challenges.

\subsection{eIM (eSIM IoT Remote Manager) Configuration}

The eSIM IoT Remote Manager is responsible for remote Profile State Management Operations on a single IoT Device or a fleet of IoT Devices. The eIM can either be a stand-alone component or a component of a higher-level functional system (e.g. device management platform).

To use the service of the eIM, the associated eIM Configuration Data shall be loaded into the eUICC. Only one eIM Configuration Data is allowed to be configured in the eUICC simultaneously~\cite{SGP31}.

\subsubsection{Add eIM Configuration Data via IPA}

The following procedure describes adding eIM Configuration Data to the eUICC when no eIM is associated within the eUICC.

Start Conditions:
\begin{enumerate}
    \item[a. ] No eIM is associated within the eUICC
\end{enumerate}

Procedure:
\begin{enumerate}
    \item[1. ] The IPA sends the eIM Configuration Operation, including the eIM Configuration Data to the eUICC.
    \item[2. ] The eUICC checks if an Associated eIM exists.
    \begin{enumerate}
        \item[a. ] If no eIM is associated, the eUICC executes the eIM Configuration Operation, else
        \item[b. ] the eUICC aborts the procedure.
    \end{enumerate}
    \item[3. ] The IPA retrieves the result of the eIM Configuration Operation from the eUICC
\end{enumerate}
End Conditions:
\begin{enumerate}
    \item[a. ]  The eIM Configuration Data of the Associated eIM is stored in the eUICC.
\end{enumerate}

\subsubsection{eIM Configuration via eIM}
The following procedure describes the eIM Configuration Operation process when an eIM is associated within the eUICC.
Start Conditions:
\begin{enumerate}
    \item[a. ] An eIM is associated with an eUICC.
\end{enumerate}
Procedure:
\begin{enumerate}
    \item[1. ] The eIM prepares and signs an eIM Configuration Operation and sends it to the IPA.
    \item[2. ] The IPA sends the signed eIM Configuration Operation to the eUICC.
    \item[3. ] The eUICC verifies that the eIM Configuration Operation is signed by an eIM that is configured in the eUICC as an Associated eIM.
    \begin{enumerate}
        \item[a. ] If the verification is successful, the eUICC processes the eIM Configuration Operation else
        \item[b. ] the eUICC aborts the procedure.
    \end{enumerate}
    \item[4. ]  The IPA retrieves the signed result from the eUICC.
    \item[5. ]  The IPA includes the signed result from the eUICC into a response to the eIM.
\end{enumerate}
End Conditions:
\begin{enumerate}
    \item[a. ] The requested eIM Configuration Data is stored in or removed from the eUICC.
\end{enumerate}

\subsubsection{Remove eIM Configuration Data from the eUICC}
\begin{enumerate}
    \item[1. ] eIM Configuration performed by the EUM
    The EUM performs the loading of the eIM Configuration Data into the eUICC during the eUICC manufacturing process.
    \item[2. ] eIM Configuration Performed in the IoT Device Production
    A production tool communicates with the IoT Device and establishes a secure link to the IPA to trigger eIM Configuration and to provide the eIM Configuration Data. The IPA transfers the eIM Configuration Operations and corresponding results to/from the eUICC.
    \item[3. ] eIM Configuration Performed in the Field by a Backend System
    The following procedure describes how to completely remove all eIM Configuration Data from the eUICC.
    
    Start Conditions:
    \begin{enumerate}
        \item[a. ] An eIM is associated within the eUICC
    \end{enumerate}
    Procedure:
    \begin{enumerate}
        \item[1. ] The IPA sends the eIM Configuration Data removal operation to the eUICC.
        \item[2. ] The eUICC executes the operation and removes all available eIM Configuration Data stored in it.
        \item[3. ] The IPA retrieves the result of the operation from the eUICC.
    \end{enumerate}
    End Conditions:
    \begin{enumerate}
        \item[a. ] The eIM Configuration Data is completely removed from the eUICC.
        \item[b. ] The eUICC is not associated with any eIM anymore.
    \end{enumerate}
\end{enumerate}

\subsection{Profile Downloading}

The IoT device must support at least one of the following three Profile Download mechanisms ~\cite{SGP31}:
\begin{itemize}
    \item Profile Download from default SM-DP+
    \item Profile Download with Activation Code
    \item Profile Download via SM-DS
\end{itemize}

There are two kinds of profile download procedures: direct and indirect download. The direct profile download requests the IoT device to directly establish the connection with the RSP server (e.g., SM-DP+) to download the profiles. The indirect profile download, on the other hand, will need help from the eIM to download the Profile from the RSP server and transfer the Profile to the IoT device. Considering the direct and indirect ways of download, the Procedures for IoT devices can be further classified into the following five approaches:
\begin{itemize}
    \item Profile Download Triggered by eIM with Activation Code
    \item eIM Initiated Direct Profile Download with SM-DS
    \item eIM Assisted Profile Download Triggered by eIM with Activation Code
    \item Profile Download with Default SM-DP+
    \item eIM Assisted Profile Download Triggered by eIM with SM-DS
\end{itemize}
The major difference between the IoT Provisioning and the traditional Provisioning is the involvement of the eIM in the procedure of the profile downloading. Here list two examples of eIM involved Profile Downloading procedure: Profile Download Triggered by eIM with Activation Code and eIM Assisted Profile Download Triggered by eIM with Activation Code.

\subsubsection{Profile Download Triggered by eIM with Activation Code}

In the general procedure of Profile Download with Activation Code, the process starts with the end user scanning the QR code or manually input the activation code. With the eIM involved, this step is replaced with the activation code transfer from eIM to the IPA. The rest steps are the same as the original process without the eIM.

\begin{figure}[htbp]
    \centering
    \includegraphics[scale=0.5]{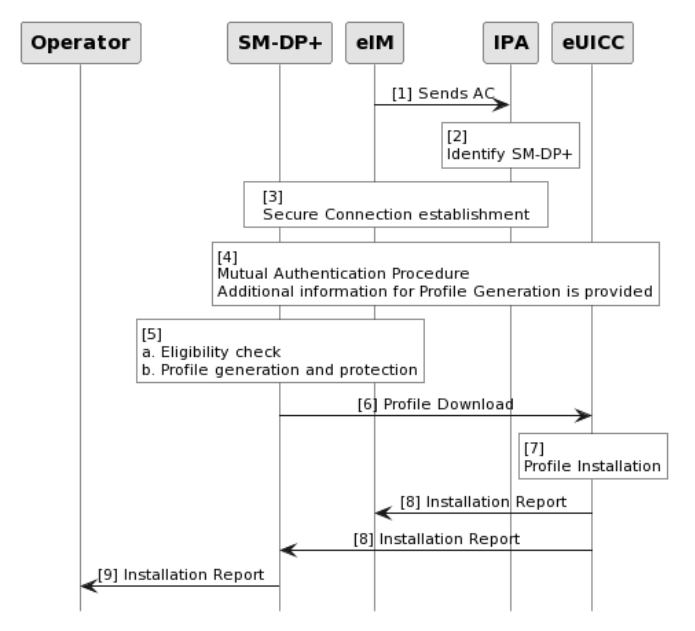}
    \caption{Procedure of Profile Download Triggered by eIM with Activation Code}
    \label{fig:IoT-ProfileDownload-eIMTriggered}
\end{figure}

\subsubsection{eIM Assisted Profile Download}

The following procedure describes the indirect Profile Download procedure between the SM-DP+ and the eUICC where the eIM assists with the Profile download. The Profile download is triggered by the eIM using an Activation Code.

In this approach, the eIM works as the proxy between the SM-DP+ and the device IPA. The eIM takes the responsibility to communicate with the external public RSP components for the device IPA and assist data exchange

Start Conditions:
\begin{enumerate}
    \item[a. ] The ordering process related to this Profile has been completed.
    \item[b. ] The Activation Code is available at the eIM.
\end{enumerate}
Procedure:
\begin{enumerate}
    \item[1. ] The secure connection between the IPA and the eIM is established via ESipa.\\
        The ESipa is a logical interface between the IPA and the eIM. The actual implementation varies depending on the connecting media and manufacturer. The ESipa is a logical interface between an eIM and an IPA. It shall provide a secure transport for the delivery of PSMO between an eIM and an IPA, unless the underlying transport provides necessary security.\\
        The format of data transported via ESipa also varies depending on different implementation. The format of data can be unified in JSON or ASN.1 or any other format that best serves the needs of underlaying protocol.
    \item[2. ] The eIM parses the Activation Code (AC) to identify the SM-DP+ address.
    \item[3. ] The eIM establishes a secure connection with the SM-DP+.
    \item[4. ] Mutual Authentication between eUICC and SM-DP+ is performed. The mutual authentication is initiated and driven by the eIM on behalf of the IPA and involves relaying authentication messages between the IoT Device and SM-DP+ including re-encoding of the messages for the two different secure connections.\\
        NOTE: The Matching Id from the AC is provided by the eIM to IPA as part of the mutual authentication exchange.
    \item[5. ] The SM-DP+ proceeds with the Profile preparation:
    \begin{enumerate}
        \item[a. ] Performs the eligibility check based on the provided eUICC and IoT Device information.
        \item[b. ] Prepare the Bound Profile Package.
    \end{enumerate}
        NOTE: The Operator owning the Profile SHALL be able to stop the Profile download at this stage.
    \item[6. ] The eIM receives the Bound Profile Package from the SM-DP+ using the secure connection with SM-DP+.
    \item[7. ] The Bound Profile Package is loaded to the eUICC:
    \begin{enumerate}
        \item[a. ] The eIM sends a request to IPA to load the Bound Profile Package to the eUICC. The request contains the Bound Profile Package and is sent using the secure connection with the IoT Device/IPA.
        \item[b. ] IPA loads the Bound Profile Package to the eUICC.
    \end{enumerate}
    \item[8. ] The Profile contained in the Bound Profile Package is installed by the eUICC.
    \item[9. ] Successful installation of the Profile is reported back to the eIM in the response to the request from the eIM. The response contains a Profile installation result Notification signed by the eUICC.
    \item[10. ] The eIM delivers the Notification to the SM-DP+ using the secure connection with SM-DP+.
    \item[11. ] The Operator is notified by the SM-DP+ about the Profile Installation
\end{enumerate}
End Conditions:
\begin{enumerate}
    \item[a. ] A Bound Profile Package has been downloaded and the Profile contained in the Bound Profile Package is installed on the eUICC. The Profile is in Disabled state.
\end{enumerate}

\begin{figure}[htbp]
    \centering
    \includegraphics[width=\linewidth]{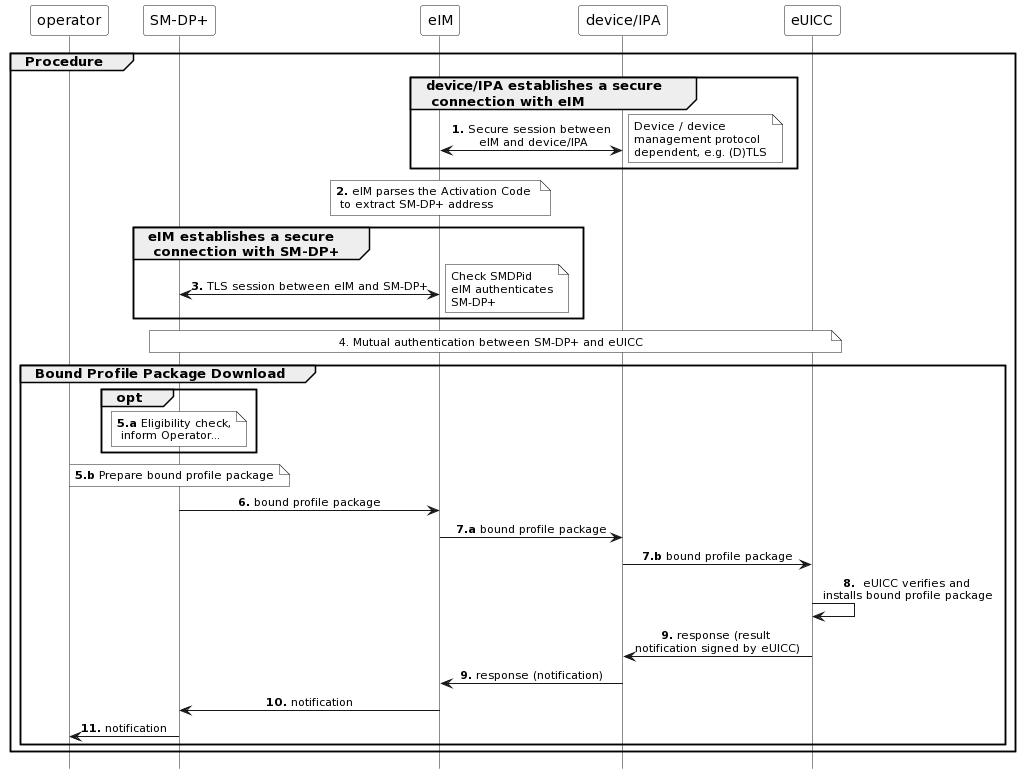}
    \caption{Procedure of eIM Assisted Profile Download}
    \label{fig:IoT-ProfileDownload-eIMAssisted}
\end{figure}
\section{conclusion}

eSIM’s remote SIM provisioning is a better alternative to traditional SIM cards and will be massively deployed in the future. It significantly simplifies provisioning or switching subscription profiles and offers greater flexibility for end users. The embedded SIM chip also saves space and reduces the weight of a device. eSIM will also contribute to developing IoT devices that can provision SIM service more efficiently and conveniently. The overall profile management difficulties of IoT devices due to the large volume of devices will also be significantly resolved.

\bibliographystyle{IEEEtran}
\bibliography{bibs/references}

\begin{thebibliography}{10}
\providecommand{\url}[1]{#1}
\csname url@samestyle\endcsname
\providecommand{\newblock}{\relax}
\providecommand{\bibinfo}[2]{#2}
\providecommand{\BIBentrySTDinterwordspacing}{\spaceskip=0pt\relax}
\providecommand{\BIBentryALTinterwordstretchfactor}{4}
\providecommand{\BIBentryALTinterwordspacing}{\spaceskip=\fontdimen2\font plus
\BIBentryALTinterwordstretchfactor\fontdimen3\font minus
  \fontdimen4\font\relax}
\providecommand{\BIBforeignlanguage}[2]{{%
\expandafter\ifx\csname l@#1\endcsname\relax
\typeout{** WARNING: IEEEtran.bst: No hyphenation pattern has been}%
\typeout{** loaded for the language `#1'. Using the pattern for}%
\typeout{** the default language instead.}%
\else
\language=\csname l@#1\endcsname
\fi
#2}}
\providecommand{\BIBdecl}{\relax}
\BIBdecl

\bibitem{GSMA-eSIM}
\BIBentryALTinterwordspacing
{GSMA}. (2023) {GSMA eSIM}. [Online]. Available:
  \url{https://www.gsma.com/esim/}
\BIBentrySTDinterwordspacing

\bibitem{GSMA-embedded-SIM-arch-1.1}
\emph{Embedded SIM Remote Provisioning Architecture}, GSMA, 2013.

\bibitem{eSIM-Whitepaper}
{GSMA}, ``{eSIM Whitepaper: The what and how of Remote SIM Provisioning},''
  {GSMA}, Tech. Rep., 3 2018.

\bibitem{SGP21}
\emph{eSIM Architecture Specification}, GSMA, 3 2022, version: SGP.21 V3.0.

\bibitem{SGP01}
\emph{Embedded SIM Remote Provisioning Architecture}, GSMA, 11 2022, version:
  SGP.01 V4.3.

\bibitem{esim-mep}
\BIBentryALTinterwordspacing
{Android Open Source Project}. (2023) {Multiple Enabled Profiles}. [Online].
  Available: \url{https://source.android.com/docs/core/connect/esim-mep}
\BIBentrySTDinterwordspacing

\bibitem{SGP31}
\emph{eSIM IoT Architecture and Requirement Specification}, GSMA, 5 2023,
  version: SGP.31 V1.1.

\bibitem{esim-overview}
\BIBentryALTinterwordspacing
{Android Open Source Project}. (2023) {Implementing eSIM}. [Online]. Available:
  \url{https://source.android.com/docs/core/connect/esim-overview}
\BIBentrySTDinterwordspacing

\bibitem{esim-euicccardmanager}
\BIBentryALTinterwordspacing
------. (2023) {Implementing eSIM - EuiccCardManager}. [Online]. Available:
  \url{https://source.android.com/docs/core/connect/esim-overview#EuiccCardManager}
\BIBentrySTDinterwordspacing

\bibitem{SGP22}
\emph{RSP Technical Specification}, GSMA, 10 2022, version: SGP.22 V3.0.

\end{thebibliography}




\bibitem{esim-overview}
  \emph{Implementing eSIM},
  Android Open Source Project,
  \url{https://source.android.com/docs/core/connect/esim-overview},
  2023,
  \emph{Accessed: 2023-08-14}.

\bibitem{SGP22}
  GSMA,
  \emph{RSP Technical Specification},
  2022,
  \emph{October},
  \emph{Version: SGP.22 V3.0}.
\end{document}